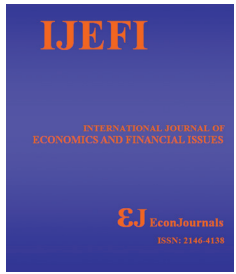

**International Journal of Economics and Financial Issues**

ISSN: 2146-4138

available at http: www.econjournals.com

International Journal of Economics and Financial Issues, 2017, 7(6), 182-191.

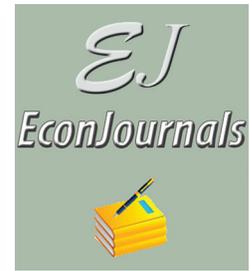

# The Determinants of Home Bias in Stock Portfolio: An Emerging and Developed Markets Study

**Mounira Chniguir[1], Mohamed Karim Kefi[2], Jamel Eddine Henchiri[3*]**

[1]Doctorante, RED-ISGG rue Jilani Lahbib 6000 Gabès, Tunisia, [2]Enseignant-Chercheur, RED-ISGG, CERI Paris, France, [3]Maitre de Conference, RED-ISGG rue Jilani Lahbib 6000, Gabès, Tunisia. *Email: jamelhenchiri@yahoo.fr

**ABSTRACT**

The objective of this paper is to measure the degree of home bias (HB) within holdings portfolio and to identify their determining factors. By following literature and an international capital asset pricing model, we have chosen quite a number of susceptible factors that impact HB. This model is, hence, estimated for 20 countries, with cross-section econometrics, between 2008 and 2013. Our results show that all countries have recorded a high level of HB in their holdings portfolio. After that, we test if the HB of the emerging markets and that of the developed markets react differently to the determining factors. The volatility of the exchange rate is statistically significant with emerging markets, while it is hardly remarkable for the developed countries. Co-variance, size, distance, language, legal framework and foreign organization stocks prevents American investors to invest abroad.

## 1. INTRODUCTION

Financial markets are characterized by quite an important volatility and a risky environment, wherefore the investor finds himself confronted to manage dynamically his portfolio. Within this context, the portfolio selection by Markowitz (1952; 1959) concentrated on this problem mainly upon the returns-risk paradigm. This theory stipulates that the inclusion of less correlated securities within a portfolio drastically reduces its risk. However, this risk may also be reduced by international diversification as geographical diversification of the portfolio across many markets over the world.

Solnik (1974), Harvey (1991), De Santis and Gérard (1997), Lewis (2000) and Arouri (2005; 2006) show that international diversification allows to downsize considerably the risk on portfolio and/or to improve its anticipated yields.

Nevertheless, despite the recommendations of the financial theory prompting the importance of international diversification, institutional investors show strong preference for national assets.

This national preference is called "home bias" (HB). Many studies focused on the factors that are significantly contributing to the existence and justification of HB.

With the recent rise of emerging markets, our objective consists in resolving the following question:

What are the benefits of international diversification?

What are the explanatory factors of HB?

We study the determinants of HB in connection with the newly emerging and developed markets both jointly and separately. First, we present the objective of our work. After that, we bring forward in the second section a brief review of the major works and latest developments concerning the topic of international diversification. The data and the methodology will be the object matter of the third section. The fourth section exposes the results of our empirical tests. At last, we shall finish by discussing our results in a fifth section, while we shall conclude in a sixth section.

## 2. A SUMMARY OF LITERATURE

The international diversification seeks to improve the performance of portfolio.

With Grubel (1968), Levy and Sarnat (1970) and Solnik (1974), the international allotment of investments constitutes the best way to ameliorate the performance of portfolios. Their idea was to derive the efficient portfolio out of the international stock-markets.

Grubel (1968) supposes that international portfolio diversification represents a mean of gain through making out international relations that differ from those coming out from the traditional trade. He tested the stock-market indexes of 11 countries over the period spanning from January 1959 to December 1966 in order to examine the advantages of international diversification for American investors. He proved that the international diversification between the assets allowed investors to reach either the highest rates of return on investment, or the lowest variance of their portfolios.

Extending from this contribution, Levy and Sarnat (1970) have documented low coefficients of correlation between the returns on assets and concluded that the gain from international diversification was substantial.

French and Poterba (1991) demonstrate the strong proclivity of American, British and Japanese investors to hold national securities.

Tesar and Werner (1995) observed the same phenomenon in a dynamic way from 1970 to 1990 with a weak addition of foreign securities in the portfolios of German, American, British, Canadian and Japanese investments.

Liljeblom et al. (1997) investigated about the benefits of international diversification concerning Nordic European countries. They analyze, over the period between 1974 and 1993, the monthly MSCI returns of 18 national capital markets. They found substantial gains for all Nordic countries with the exception of Denmark.

Likewise and Lapp (2001) analyzed the benefits of international diversification for the German investor. This study, on a sample of 18 countries throughout a period of 10 years, from 1988 to 1997, showed that a diversified portfolio was much more profitable than a purely domestic portfolio without taking into account the degree of aversion to risk. However, Schroder (2002) showed that the strategy of international diversification permits to give additional and significant returns for all investors belonging to his sample (British, French, and German investors) with the exception of the American investor.

Li et al. (2003) studied the behavior of the monthly G7 indexes with 8 newly emerging markets from Latin America and Asia from January 1976 to December 1999. They concluded that the anticipated gains from international diversification remain substantial for investors in the American stock market wherein there are no constraints of short sale on the newly emerging markets. They remarked a reduction of financial integration level on international markets, but this does not annul the anticipated gains from the international diversification.

Arouri (2005) elaborated an extension of the generalized autoregressive conditional heteroscedasticity multi-varied model in order to derivate a measurement of the predicted gains from the international diversification of the portfolio made from the stock markets of the G7 countries and the international markets, in the period from February 1970 to May 2003. The results show that the benefits of international diversification are statistically and economically significant for all financial markets except those of France and UK. The gains depend upon the low correlation between return of the different securities invested in the markets. Any increase of such correlation may be responsible for the decreases of the predicted benefits of international diversification, which results from the rise of the level of integration of financial markets. Consequently, the gains of diversification are linked with the level of integration or segmentation of financial market.

In order to better profit from the strategy of international diversification it was showed that investors go towards the developed markets. Nevertheless, with the turning of the 1980s and 1990s, many studies assess the significance offered by the emergent stock markets. Beforehand, these markets had to be able to manage important potentials of diversification in comparison with the markets of developed countries which have a tendency to be more and more integrated with each other. These changes can be observed by the low correlation of the newly emerging markets with one another and with the rest of the world.

Odier et al. (1995) examined the characteristics of returns for the newly emerging markets in comparison with the developed markets. They recorded an important increase of returns in the newly emerging markets, but these returns were associated with high levels of risk. They concluded that the correlation between returns within the newly emerging markets and the international index of the developed markets was 0.31, which means that the gains anticipated from the international diversification on the newly emerging markets is much more beneficial than investing uniquely on the developed markets.

Bellalah and Fadhlaoui (2006) gave evidence of the importance of the newly emerging markets for the international diversification by measuring the relations of interdependence between the developed markets and the newly emerging ones. The benefits of international diversification resides first in the reduction of the risk, and second, in the improvement of performance. However, in spite of the theoretical and empirical evidences about the additional gains of international diversification, investors tend to allot part of their wealth in foreign securities weaker than that which they invest in domestic securities. Investors hardly diversify their portfolios and prefer to hold local securities.

Moreover, the preference of domestic assets is considered to be a normal behavior as Gorman and Jorgen (2002) showed. Their study, included the investors of 5 greatest countries in the world, showed that the sharp ratio for the optimal portfolio is typically below the Sharpe ratio for the portfolio totally domestic. Therefore, the totally allocation portfolio is much more profitable than the one with optimal portfolio. To sum up, they conclude that domestically-oriented investors are not "irrational" and that the benefits of international investment are hard to attain.

In the same way, Cambell (2006) suggests that investors tend to prefer national securities. He conducts his study such as French and Porterba (1991), Cooper and Kaplanis (1994) in order to show that domestic investors have a very high level of "Household" comparatively with foreign investors. Besides, investors exhibit a preference for regional companies against non-regional firms.

HB can be explained mainly by the existence of barriers against the flux of capital like the costs of trade, the withholdings tax, the political risk, the failure in parity of the purchasing power, informational asymmetries and the constraints on short sales. Hence, it clearly appears that there are numerous factors which push international under-diversification.

Within this framework, Mishra (2008) fixed a given set of factors to explain national preference relying upon the economic theory. Among these factors, he used the currency exchange risk, distance, language, the joint-variance, the legal framework the cost of transaction and other informational variables. According to certain studies, the volatility of the exchange rate constitutes a brake on the growth of international investment. Within this purview, the exchange risk constitutes an important variable that has to be taken into consideration in every strategy of international investment and more particularly within the newly emerging markets. At this level, we equally have to emphasize the fact that the occurrence of financial crises in some markets is translated by important devaluations of the currency in which the securities are made out. Hence, a positive link is made up between the volatility of the exchange rate and the HB.

Bin et al. (2003) showed that the foreign exchange rate Giurda and Tzaralis (2004) and the interest rate were the principal determinants of assessment of development results assessment of the developed countries (Australia, Japan, and Europe) and those of the newly-emerging countries (China, Chile, Korea, Mexico and South Africa).

Within this context, a great deal of theoretical and empirical corpus has been conducted to explain the HB taking into account diverse frictions in the financial markets. Within this framework, taxes appear to have a strong effect on the demand of foreign assets. In fact, the inverse relation between the tax and the demand of foreign assets indicates that investment in foreign asset diminishes as tax increases.

From this point of view, Stulz (1981) concluded that tax may be at the origin of the segmentation of financial markets, which discourages the international investment and, consequently, increases the decision of the investor to prefer the local market. He proved that certain assets can only be possessed by domestic investors because their importance in terms of diversification does not compensate the cost in terms of foreign investment taxes. Hence, this result shows that the tax-factor contributes significantly to the explanation of HB.

The inverse relation between the tax and the demand of foreign assets indicates that the investment in foreign assets diminishes as the tax increases. Thence, drawing upon Black (1974), we should say that the heterogeneity of taxes explains the elucidation of financial assets and particularly the preference of investors for national assets.

Amadi (2004) supported the idea that common language, distance, as well as immigrants equally influence the strategy of foreign investment. Such costs may hinder the entrance to a foreign market. In this perspective, investors move typically over markets that have the same language, culture or other domestic notions Grimblat and Kelaharju (2001).

Gordon and Hines (2002) point out that the inclusion of taxes in a standard portfolio model generates a forecast that investors tend to prefer these stocks where they face the relatively favorable tax regime compared to other investors. However, the estimates taking into account the tax do not always offer a favorable explanation for the preference observed in the portfolio by indicating that there are omissions in the models studied. Otherwise, there is the possibility of tax evasion on the income of foreign securities by using foreign financial intermediaries.

From this perspective, Gordon and Hines (2002) concluded that when "capital flight" is an important empirical phenomenon, domestic investors will appear to be foreign, so the portfolios observed should have a foreign bias rather than domestic bias, which is inconsistent with the reported proposals.

In tune with the recent study, Berkel (2004) examined the institutional determinants of international portfolios of assets. In particular, the results of estimation indicate that the asymmetries of information, approximated by geographical proximity, the common language or a common colonial background have an enormous effect upon the composition of international portfolios of assets. Thus, we are going to use the distance between the country of origin and the country of destination as a proxy variable to replace the cost of information needed to the acquisition of foreign assets. In this context, investors prefer to make their investments within the country that has the same language in use as their native countries, which helps facilitate the financial analysis of organizations.

Moreover, the size of a market improves its liquidity and its capacity to mobilize funds and diversify risks. This means that American investors are going to orient themselves towards the markets that have an important share within the global market. As a consequence, the increase of the country size favors the strategy of international diversification with this country. In fact, if the number of stocks listed in the domestic market are important, the American investors decrease their foreign investments.

**Figure 1:** Conceptual model

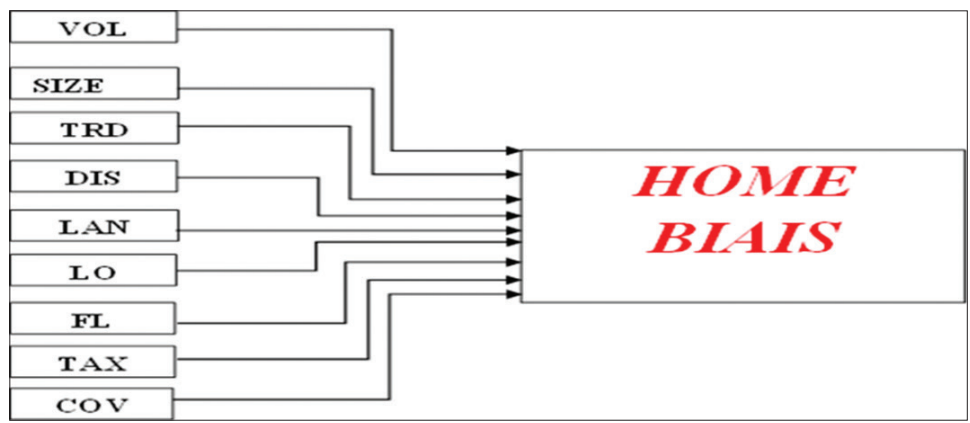

A good openness to the external trade gains the trust of international investors and attracts more foreign funds. The more the number of foreign companies increases within the American market, the more the American investors would have better information and vision about the markets of those countries, which would facilitate the holding of foreign assets while those countries would gain from the strategy of international diversification.

Our objective is to test the hypothesis that exchange rate volatility, size, transaction, distance, language, legal origin, foreign organizations' share, covariance-diversification, as well as tax, influence the phenomenon of HB for the American investor.

The conceptual model of this research is presented in Figure 1.

## 3. RESEARCH METHODOLOGY

The regional classification of our sample of 20 countries is displayed in the tables in annexes. This classification is based on the division of the geographical area. Our research spans over a period of 6 successive years, from 2008 to 2013.

The starting point of our empirical approach consists essentially in measuring the issue of HB. This descriptive analysis verifies the under-weightings of foreign assets within the portfolio for the majority of countries during the studied period. We use, as data, the total cross-border holdings of assets on one side, while on the other, the part of invested assets at national level. The data are excerpted from World Exchange Federation. Besides, it is indispensable to note that the investors of the sample-countries effectively make the exchange of financial assets with the American market and vice-versa.

In this context, we test if the HB is observed for the countries within our sample during the indicated period. This means that the investors significantly hold portfolios of assets which are largely biased towards domestic securities. Likewise, we are going to verify if we invest a much more important proportion in the domestic market than that being suggested by the theory. Indeed, after measuring the countries global HB, we analyze the HB in a bilateral way. We focus solely on the study of foreign shares-holding by American investors. Then, we have to test the determinants of HBs based upon an econometric model.

## 4. IMPLEMENTATION AND RESULTS

As Mishra (2008), the explanatory variables such as the overall measurement of HB of a given country is a relationship between the real relative share of assets invested in the domestic market and the aggregate assets of the country.

Formally:

HB = Real share of domestic holding/aggregate asset

With:

Aggregate asset = domestic asset + foreign asset + investment fund

After analyzed the aggregate holdings of countries foreign assets and in order to better locating the distribution of HB, we intend to implement a bilateral analysis of the bias as Mishra (2008) who studied the holding of Australian investors.

Our point of focus is to study holdings portfolios relative to American investors, who are concerned mainly with American market. So, we are going similarly to quantify the volume of the HB of a given country i towards country j as the relative difference between the weight of domestic foreign portfolios and the optimal ones. In other words, we have to prove the existence of HB, in regard to American portfolios, through the comparison of the actual proportion of foreign assets in American investors' portfolios coming from the selected countries of the sample with that predicted by the global market.

Formally:

$$HB_{ij} = 1 - \frac{I_i^j}{I_i^*}$$

The holding of foreign assets within the total share portfolio of American investors is attained *via* the relationship between their holdings of foreign shares and their aggregate holdings of shares (foreign and domestic).

Our econometric model of international capital asset pricing model made these factors obvious and discernible. Indeed, the main point within this model is to present and discuss the pertinence of the variables to be the same self-factors determining the issue of HB. We are going to verify by measuring the potential causes of this bias between the financial theory and the economic reality.

In fact, our model encompasses the exchange risk variable with the other variables that measure the frictions in the financial market, and it is tested through the ordinary least squares (OLS) econometric method using cross-country data. So, the estimated basic statistical model is formulated as follows:

$$HB_{ij} = \alpha_1(VOL) + \alpha_2(COV) + \alpha_3(SIZE) + \alpha_4(LAN) + \alpha_5(LO) + \alpha_6(FL) + \alpha_7(TRD) + \alpha_8(DIS) + \alpha_9(TAX) + \varepsilon_{ij}$$

With:
$HB_{ij}$: The dependent variable which indicates the bilateral measurement of HB.
The volatility of the exchange rate (VOL): Measured by the standard deviation of monthly change of the bilateral exchange rate over the specific period of study.
Diversification-covariance (COV): Indicates the covariance between the American market's returns and the country of destination.
Language (LAN): A dummy variable that can take two values: 1, if both countries either have a common language or a common colonial background, and 0, if it is not the case. This variable indicates the legal system in use within the country.
Foreign listing (FL): Is the number of foreign firms listed on total number of firms listed in domestic market. This measure is credited to Mishra (2008) and Amadi (2004).
Trade (TRD): Measures the average of imports and exports normalized by the destination country's gross domestic product (GDP). This variable is given under the following formula: 0.5* (exports + imports)/GDP.
Distance (DIS): Indicates the physical distance in kilometers between the country of origin (the USA) and the country of destination. This variable is estimated using the logarithm.
The tax (TAX): Indicates the set of taxations imposed by the country.
$\varepsilon_{ij}$: Indicates the margin of error.

The real holdings of portfolio are identified by using the date of international investment locations. Indeed, the following Table 1 displays the different elements to be considered in the measurement of HB for the year 2012.

The figures permit some observations: Besides, it is highly interesting to remark that all countries record a high level of HB. This implies that investors of these countries significantly hold share portfolios that are biased towards national securities. In other words, investors prefer to invest within the national market, while they ignore the potential gains of international diversification strategy.

In 2012, the least elevated value of HB concerning portfolios was registered in the UK where only 56.10% of the total volume of share portfolios was invested in domestic stock. The most elevated value was registered in Columbia, India and Turkey where we find

**Table 1: HB in share portfolios (December 2012)**

| Country | Aggregate assets | Aggregate domestic assets | HB |
|---|---|---|---|
| Australia | 672388.5 | 636089.8 | 0.946015287 |
| Brazil | 165275.6 | 164816.1 | 0.997219795 |
| Canada | 900493.1 | 898851 | 0.998176443 |
| Columbia | 9418.9 | 9418.9 | 1 |
| China | 392772.4 | 387003 | 0.985311086 |
| Egypt | 26241.1 | 26239.4 | 0.999935216 |
| Spain | 1566107.1 | 1546869.7 | 0.987716421 |
| South Africa | 201779.1 | 145687 | 0.722012339 |
| Germany | 1915304.5 | 1749657.5 | 0.913514013 |
| India | 473671.2 | 473671.2 | 1 |
| Japan | 4679557.8 | 4622614.8 | 0.987831542 |
| Malaysia | 51601.4 | 50449.3 | 0.977673086 |
| Mexico | 56682.8 | 53392.4 | 0.941950644 |
| Poland | 30421.5 | 29290.1 | 0.962809197 |
| France | 2906208.2 | 2776739.5 | 0.955450989 |
| United Kingdom | 5677721 | 3185654.2 | 0.561079736 |
| Thailand | 95645.8 | 95637.1 | 0.999910084 |
| Turkey | 199187.8 | 199187.8 | 1 |
| USA | 28553196.4 | 25113766.3 | 0.879543079 |
| Italy | 1293682.1 | 1216676.3 | 0.940475484 |

HB: Home bias

**Table 2: Bilateral bias in the share portfolio in 2012**

| Country | Real portion of foreign shares | Optimal portion | $\dfrac{I_i^j}{I_j^*}$ | $HB_{ij}$ |
|---|---|---|---|---|
| Australia | 0.0187 | 0.71738 | 0.026 | 0.97393 |
| Brazil | 0.0477 | 0.3459 | 0.1379 | 0.86209 |
| Canada | 0.1160 | 0.350 | 0.3314 | 0.6686 |
| Columbia | 0.00081 | 0.1677 | 0.00483 | 0.99517 |
| China | 1.136 | 8.978 | 0.1265 | 0.87347 |
| Egypt | 0.0544 | 0.526 | 0.10342 | 0.89658 |
| Spain | 0.0083 | 0.2617 | 0.03171 | 0.96829 |
| South Africa | 0.0017 | 0.8742 | 0.001956 | 0.99804 |
| Germany | 0.0255 | 0.4442 | 0.05733 | 0.94266 |
| India | 0.0051 | 0.7037 | 0.007218 | 0.99278 |
| Japan | 0.335 | 1.418 | 0.23624 | 0.76375 |
| Malaysia | 0.01068 | 0.282 | 0.0378 | 0.962127 |
| Mexico | 0.00128 | 0.3459 | 0.00370 | 0.9963 |
| Poland | 0.00048 | 0.1229 | 0.00387 | 0.99612 |
| France | 0.0364 | 0.3614 | 0.10072 | 0.8993 |
| United Kingdom | 1.475 | 6.351 | 0.23224 | 0.7977 |
| Thailand | 0.04727 | 0.7739 | 0.061073 | 0.9389 |
| Turkey | 0.0185 | 0.1326 | 0.139215 | 0.86078 |
| Italy | 0.02101 | 0.327 | 0.06425 | 0.93575 |

that all investments in portfolio were domestic. It seems that HB in shares holding diminished within developed countries' markets (US, France, Germany), whereas the markets of the emerging ones (Brazil, UK, Egypt) record a much higher level of HB. In this respect, it is proven that the developed countries markets offer the best opportunities to diversify the share portfolios at international level, which means that the correlations between the financial markets of these countries are weak.

Now, we present a bilateral analysis of the bias, after analyzed the holdings of aggregate foreign assets of countries. Indeed, the Table 2 presents the components of bilateral bias in 2012, by computing this bias between the American market and the other countries of the sample.

We observe that the American portfolio is roughly composed of 85% of American stock. As we have already mentioned, American investors have strong preferences for domestic shares; they tend to invest less internationally than what is theoretically predicted, and the real foreign participation is smaller than the optimal portion of international assets within investors' portfolios.

**Table 3: Results of regression of the global econometric model (2013)**

| variables | Coefficient | Statistic-t |
|---|---|---|
| VOL | 0.18 | (1.69) |
| COV | 0.00843 | (3.91)*** |
| SIZE | −3.1034 | (−2.254)** |
| DIS | 2.4E-07 | (1.68)* |
| LAN | −1.041 | (−0.067)* |
| LO | −0.249 | (−0.00)*** |
| TAX | 0.076 | (0.015)** |
| TRD | −1.65E-07 | (1.048)** |
| FL | −0.59 | (−8.95)* |
| CONST | 1.867 | (1.419) |

$R^2$=0.27. Number of observation: 171
*Means threshold at 1%, **means threshold at 5%, ***means threshold at 10%

**Table 4: Results of regression of the basic econometric model: Emerging countries**

| variables | Coefficient | Statistic-t |
|---|---|---|
| VOL | 0.011 | (0.016)*** |
| COV | 0.0084 | (3.91)*** |
| SIZE | −0.21 | (−0.13) |
| DIS | 0.17 | (2.71)* |
| LAN | −0.08 | (−3.77)* |
| LO | −0.06 | (−3.63)* |
| TAX | 0.003 | (2.14) |
| TRD | −0.005 | (−2.4)** |
| FL | −0.0001 | (−0.213) |
| CONST | 0.9 | (16.72)* |

$R^2$=0.59
Number of observation: 99
Source: Our calculations/figures, *means threshold at 1%, **means threshold at 5%, ***means threshold at 10%

**Table 5: Results of regression of the basic econometric model: Developed countries**

| variables | Coefficient | Statistic-t |
|---|---|---|
| VOL | 0.0001 | (1.6) |
| COV | 0.0072 | (3.35)*** |
| SIZE | −0.43 | (−0.07)*** |
| DIS | 0.42 | (5.35)* |
| LAN | −0.11 | (−6.65)* |
| LO | −0.09 | (−4.52) |
| TAX | 0.005 | (2.4) |
| TRD | −0.008 | (−3.91) |
| FL | −0.0007 | (−0.149) |
| CONST | 0.8 | (10.3) |

$R^2$=0.54
Number of observation: 72
Source: Our calculations/figures, *means threshold at 1%, **means threshold at 5%, ***means threshold at 10%

A multi-factorial econometric model is used to analyze the factors that affect the choice of investors' financial assets. We estimate this model relying upon the OLS model using of cross-section data for the period from 2008 to 2013.

The results for the year 2013 are presented in the Table 3.

The analysis of Table 3 allows us to deduce that the variable of the volatility of the exchange rate is positively correlated with HB phenomenon at (0.18) for the year 2013. The positive figure obtained for the exchange risk is conforming to economic intuition. This means that the volatility of the exchange rate generally reflects the uncertainties towards the economic situation of the country and towards the efficiency of the predicted shares. A strong volatility of the exchange rate encourages the American investors to invest in the domestic market, and consequently, it represents a barrier against the international movements of capitals, which sustains the results attained by Mishra (2008) on the Australian market. The exchange risk is a strong factor to explain national preference.

Nevertheless, our results could not offer enough information to determine whether the considered factors-seemingly the cause of HB-had specifically been at a certain sub-sample. That's why, we have classified our sample-countries into developed (Australia, Canada, Germany, Japan, France, Spain, UK) and newly-emerging ones (Brazil, Egypt, Poland, Thailand, Turkey, China). In the light of this classification, we have estimated the determining factors of HB for every category aside. Therefore, the following Tables 4 and 5 display the results of this estimation.

More precisely, when we separately examine the developed countries and the newly-emerging cases, we find that the volatility of the exchange rate is generally of low importance for the developed countries' markets during this period, while it is quite important at a threshold of 1% for the years 2008, 2009 and 2011, and at a threshold of 10% for the years 2010, 2012 and 2013 concerning the emergent markets. Hence, the volatility of the exchange rate encumbers American investors from embarking on the markets of the newly-emerging economies.

The covariance factor obtained is positive at (0.00843) for the year 2013, and it is strongly important at a threshold of 10% for almost the whole period of study. As for the emerging countries, the coefficient of this variable is positive and interesting over the whole period of study at a threshold of 10%. This indicates that there is a strong correlation between the return of the American market, which consequently means that domestic assets are strongly weighed within the portfolios of American investors.

Yet, for all the developed countries, the co-variance is positive, but not strong, except for the years 2012 and 2013. So, we conclude that the American market is weakly correlated with the markets of developed countries, which facilitates the acquisition of foreign assets by these countries and, consequently, the American investors will benefit from the strategy of international diversification. We find that American investors prefer invest in the markets of these countries, even though the emerging countries seem to be more and more attractive.

## 5. THE DETERMINANTS OF HB

Moreover, we have used the distance between the country of origin and the country of destination as a proxy-variable to replace the costs of information. Within this frame, we find an interesting result that the coefficient of this variable (distance) is quite significant, which means that the distance between the financial centers is important in determining the choice of investors of their assets. Nevertheless, this coefficient is positive, which gives credence to the previously mentioned hypothesis. This indicates that the holding of foreign assets increases when the distance between the two countries increases. This can be explained by the fact that the length of distance between countries creates much higher costs to get to information.

Within this same purview, the size of the country influences HB. The coefficient of this variable is negative at (−4.89) and (−3.51), and quite significant at a threshold of 1% for the years 2008 and 2010, and a threshold of a 5% for the years 2009 and 2013. As the country stocks increases within the global market, the holding of domestic assets decreases. Hence, investors prefer to go towards the developed markets more than the emergent ones to profit from the strategy of international diversification.

We focus now on the significance of mute variables selected in this research which manifest essentially in language and legal framework. These variables indicate the cultural, linguistic and historical links between the American market and those markets making the sample of our study.

The model show that the coefficients of these variables are negative (−0.384) and (−1.15) for the years 208 and 2012; and (−0.44) and (−1.04) for the years 2011 and 2013. These coefficients are equally important at a threshold of 5% for the first range of values, and at a threshold of 1% for the others. Besides, the same result has been remarked concerning the estimation of these variables for both categories of countries (developed and emerging). These results show us that, in the case where the country of destination and the United States have the same language in use, American investors additionally increase their investments towards that country. Likewise, the same legal framework permits the countries to reinforce diversification and, hence, to diminish the impact of HB.

The results which we have attained emphasize the importance of informational functions as being the fundamental determinants of holdings of international share portfolios. In other words, both variables (language and legal origin) have a prominent role in the reduction of HB phenomenon, and they also have some important effects on the extent of diversification.

About the role that trade plays in the process of HB. The resulting figures are negative at the values of (−0.00609) and (−1.27) for the years 2008 and 2012, which goes in tune with the economic intuition. This variable is important at a threshold of 10% equally for all the markets of our sample of study as for the emergent and developed markets when treated separately. The classical trade theory emphasizes the positive relation between the trade openings and the international diversification strategy. This indicates that American investors would head towards the countries that have a respectable level of development concerning their financial markets. The effect seems much more important for the country of destination. Therefore, the more developed the financial market of the country of destination is, the more advantageous the opportunities of investment are, that's wherefore American investors would be inclined, in advance, to hold the portions of portfolio in such economy.

Concerning tax, that this variable is positively correlated with national preference (0.044) and (0.076) for the years 2010 and 2013. This means that the increase of tax rates in the country of destination leads American investors to invest less at international scale. In this context, the obtained coefficients are most often statistically important at a threshold of 5% equally for all the sample countries together as for all the emergent and developed markets when taken independently. From this perspective, we need to note that the increase of tax rate prevents American investors from holding foreign assets and, hence, they prefer domestic assets.

Furthermore, the larger the share of foreign companies listed on the domestic market, the greater their visibility by local investors is good so the access of foreign equities by domestic investors is strong. This implies that as if the number of these companies increased in the US market, most US investors have the opportunity to gather information about the markets of countries of origin of these firms, facilitating the holding of foreign assets. Therefore, this result helps to reduce the magnitude of the HB.

The estimated results of our sample demonstrate that this variable constitutes an unfavorable determinant of HB, in such way that it shows negative coefficients (−0.4) for the year 2008. Similarly, for the remaining period of study, we observe that this variable seems to be highly significant at a threshold of 1%.

Generally, the results of estimation for the year 2008 show a median ($R^2$) at the value of 58%, which means that the model has a moderate explanatory power. In general, the global model is significant during the period of study, as well as the ones on emerging markets and developed markets.

## 6. CONCLUSION

Our analysis means that the HB persists on markets all the time and at same levels in spite of the process of financial integration followed by the all countries. The results of our econometric model succeeded to demonstrate its applicability on financial reality. In this respect, we have deduced that numerous factors can explain the allocation and allotment of financial assets.

The results have allowed us to draw an important conclusion that the volatility of the exchange rate is statistically significant for the emerging markets, while it is not quite mentionable for the developed countries. This means that currency prevents American investors from investing in those countries, which is the case for both variables of co-variance, size, distance, language, legal framework and foreign organization stocks. Another conclusion drawn from this estimation is that American investors prefer to overweigh their share portfolios by the assets of foreign countries comparatively with the emerging ones.

Moreover, there will be an interesting effect of investors' behavior in the matter of choice of assets. It seems quite important, then, to identify the impact of the behavioral approach upon the issue of HB. In this respect, an interesting track for such kind of study would be to include the behavioral factors in the explanation of HB, which will be the object of our for future researches.

## 7. ACKNOWLEDGMENTS

We are grateful for the helpful comments of all the CSIFA conference referees including those from RED-ISGG Research unit and especially the support of Prof. makram Belallah. Of course, they are not responsible for the remaining errors.

# ANNEXES

**Home bias in share portfolios (2008–2013)**

| Countries | 2008 | 2009 | 2010 | 2011 | 2012 | 2013 |
|---|---|---|---|---|---|---|
| Australia | 0.987158 | 0.984924 | 0.986333 | 0.980495 | 0.94601 | 0.9591278 |
| Brazil | 0.9937897 | 0.99477108 | 0.999411 | 0.9994836 | 0.997219 | 0.998343 |
| Canada | 0.97162904 | 0.9731054 | 0.999037 | 0.998833 | 0.998176 | 0.9932199 |
| Columbia | - | - | 1 | 1 | 1 | 1 |
| China | 0.93731195 | 0.95998830 | 0.9791862 | 0.9887764 | 0.985311 | 0.980963 |
| Egypt | - | - | - | - | 0.999935 | 1 |
| Spain | - | 0.99342152 | 0.9949510 | 0.9932745 | 0.987716 | 0.99078 |
| South Africa | 0.7256966 | 0.6789498 | 0.699519 | 0.6971232 | 0.722012 | 0.7316826 |
| Germany | 0.9052322 | 0.915937 | 0.924243 | 0.9111031 | 0.913514 | 0.9072793 |
| India | 0.593932 | 0.6520648 | 0.9996170 | 0.999863 | 1 | 0.9998890 |
| Japan | 0.995282 | 0.9887451 | 0.9925307 | 0.9883832 | 0.987831 | 0.9918181 |
| Malaysia | 0.987659 | 0.9908848 | 0.9840868 | 0.9812221 | 0.977673 | 0.9834732 |
| Mexico | 0.9633548 | 0.839045 | 0.924416 | 0.961708 | 0.941950 | 0.84123405 |
| Poland | 0.979100 | 0.9929844 | 0.9356287 | 0.9740124 | 0.962809 | 0.962619 |
| France | 0.9825534 | 0.9835264 | 0.9868904 | 0.9807607 | 0.955450 | 0.9949537 |
| UK | 0.4079382 | 0.4701182 | 0.5937629 | 0.568790 | 0.561079 | 0.565737 |
| Thailand | 0.999769 | 0.9997941 | 0.9999746 | 0.9998685 | 0.999910 | 0.9998969 |
| Turkey | 1 | 1 | 1 | 1 | 1 | 1 |
| USA | 0.9022421 | 0.9012929 | 0.89370588 | 0.882558 | 0.879543 | 0.874329 |
| Italy | 0.9459445 | 0.908333 | 0.86298210 | 0.9022199 | 0.940475 | 0.94489556 |

**Bilateral bias in the share portfolio (2008–2013)**

| Countries | 2008 | 2009 | 2010 | 2011 | 2012 | 2013 |
|---|---|---|---|---|---|---|
| Australia | 0.98295 | 0.98102 | 0.97977 | 0.9776 | 0.97393 | 0.94063 |
| Brazil | 0.99619 | 0.99918 | 0.99917 | 0.999941 | 0.86209 | 0.99909 |
| Canada | 0.94405 | 0.9476 | 0.94627 | 0.94712 | 0.66857 | 0.93180 |
| Columbia | 0.99785 | 0.99607 | 0.99842 | 0.99685 | 0.99516 | 0.99287 |
| China | 0.99815 | 0.998927 | 0.99894 | 0.99893 | 0.87346 | 0.99305 |
| Egypt | 0.9956 | 0.9945 | 0.9965 | 0.998 | 0.89657 | 0.9998 |
| Spain | 0.98922 | 0.971881 | 0.981525 | 0.962726 | 0.96828 | 0.96539 |
| Afrique de sud | 0.99840 | 0.99823 | 0.99866 | 0.998450 | 0.99804 | 0.99729 |
| Allemagne | 0.946017 | 0.93335 | 0.93298 | 0.941720 | 0.94266 | 0.90614 |
| India | - | - | - | 0.99089 | 0.99278 | 0.99461 |
| Japan | 0.3726 | 0.37264 | 0.35228 | 0.434993 | 0.76375 | 0.55497 |
| Malaysia | 0.94484 | 0.92175 | 0.95661 | 0.94063 | 0.962127 | 0.94809 |
| Mexico | - | - | 0.96162 | 0.99864 | 0.99629 | 0.996164 |
| Poland | 0.995 | 0.9765 | 0.98506 | 0.99309 | 0.99612 | 0.99837 |
| France | 0.89553 | 0.87099 | 0.89563 | 0.84782 | 0.89927 | 0.89453 |
| UK | 0.76170 | 0.74790 | 0.761387 | 0.72417 | 0.76775 | 0.74725 |
| Thailand | 0.996 | 0.82795 | 0.98424 | 0.32744 | 0.93892 | 0.9463 |
| Turkey | 0.8756 | 0.70689 | 0.88288 | 0.95591 | 0.86078 | 0.69147 |
| Italia | 0.9536 | 0.9422 | 0.76137 | 0.94279 | 0.935749 | 0.94705 |

**Results of regression of the basic econometric model**

| variables | (08) | (09) | (10) | (11) | (12) |
|---|---|---|---|---|---|
| VOL | 0.39 (2.039)*** | 0.1336 (5.02)* | 0.1412 (17.02)* | 0.17 (2.3)** | 0.03 (1.78)*** |
| COV | 0.00725 (3.02) | 0.0049 (2.58)*** | 0.00707 (4.41) | 0.01 (1.16) | 0.0531 (3.32) |
| TAILLE | −4.893 (−3.7021) | −3.8934 (−2.945)** | −3.51096 (−3.943)* | −2.2616 (−1.74) | |
| DIS | 0.572 (8.83)* | 0.543 (7.64)* | 0.598 (9.3)** | 0.4524 (2.115)** | 0.49 (4.02) |
| LAN | −0.384 (−2.55)** | −1.539 (−6.76) | −0.2059 (−1.85)*** | −0.44 (−5.42)* | −1.155 (−0.054)** |
| OL | −0.1168 (−2.06) | −0.09 (−5.53)* | −0.1572 (−1.49) | −0.27 (−6.11)* | −1.056 (−0.044) |
| TAX | 0.039 (3.13) | 0.02 (1.71) | 0.044 (0.27)** | 0.06 (2.69) | 0.071 (0.069) |
| TRS | −0.00609 (−2.66)*** | −7.65E-07 (−1.839) | −0.003 (−5.2)* | | −1.27E-07 (−1.41) |
| FL | −0.4 (−4.96)* | −0.35 (−5.25)* | −1.539 (−6.76) | −0.4 (−3.22)* | −0.182 (−0.77) |
| CONST | 0.9208 (5.209) | 0.068 (1.917)*** | 0.182 (3.933)*** | 0.079 (3.3933)* | 0.202 (0.004)*** |
| R² | 0.579 | 0.34 | 0.5755 | 0.3 | 0.34 |
| Observation | 171 | 171 | 171 | 171 | 171 |

**Results of regression of the basic econometric model: Emerging countries**

| variables | (08) | (09) | (10) | (11) | (12) |
|---|---|---|---|---|---|
| VOL | 0.2143.(8.8E+10)* | 0.004 (1.85)* | 0.039 (2.039)*** | 0.0098 (4.887)* | 0.16 (2.34)*** |
| COV | 2.798 (3.76E+11)* | 0.0084 (3.91)*** | 0.000041 (3.65) | 0.0008 (0.3) | 0.00531 (3.32)*** |
| TAILLE | −6.537 (−1.63E+1)* | −0.06 (−1.9) | −0.032 (−1.64)* | −0.142 (−0.98) | |
| DIS | 0.0706 (4.6E+10)* | 0.37 (0.06)*** | 0.24 (0.14)* | 0.18 (2.77)* | 0.26 (5.84)* |
| LAN | −1.70E-13 (−0.8505) | −0.039 (−3.13)*** | −2.4E-07 (−1.68)* | −0.3 (−2.23) | −0.07 (−2.09) |
| OL | −0.1402 (−2.88E+11) | −0.18 (−2.77)* | −0.14 (−2.14) | −0.03 (−1.8)* | −0.06 (−2.69)* |
| TAX | 0.0083 (2.27E+11) | 0.0001 (0.52) | 2.4E-07 (1.68)* | 0.04 (2.01) | 0.0027 (3.35)*** |
| TRS | −0.068 (−1.9)*** | −0.0002 (−0.12) | −0.01 (−1.16)*** | | −0.0001 (−0.5) |
| FL | −0.54 (0.778)* | −0.007 (−4.05)* | −0.602 (−7.39) | −0.3 (−2.23)** | −0.01 (−1.1)* |
| CONST | 1.43 (2.43) | 0.125 (0.436) | 0.992 (18.32) | 0.89 (16.5) | 0.98 (17.23)* |
| $R^2$ | 0.3 | 0.45 | 0.53 | 0.48 | 0.4 |
| Observation | 99 | 99 | 99 | 99 | 99 |

**Results of regression of the basic econometric model: Developed countries**

| variables | (08) | (09) | (10) | (11) | (12) |
|---|---|---|---|---|---|
| VOL | 0.002 (0.011) | 0.001 (0.85) | 0.027 (1.073) | 0.0001 (0.3112) | 0.02 (1.68)** |
| COV | 0.002 (3.35)*** | 0.0072 (3.02) | 0.000036 (2.66) | 0.0001 (0.04) | 0.049 (2.75)*** |
| TAILLE | −0.0005 (−0.68)*** | −4.19E-07 (−3.42) | −0.049 (−2.9)*** | −0.219 (−1.46) | |
| DIS | 0.068 (0.722) | 0.41 (0.08)*** | 0.28 (0.13)** | 0.27 (6.11) | 0.4 (2.81) |
| LAN | −0.011 (−0.728) | −0.061 (−5.07) | −4.19E-07 (−3.42)*** | −0.4 (−4.96) | −0.18 (−2.75)** |
| OL | −0.001 (−0.022)** | −0.26 (−5.81)* | −0.17 (−2.3) | −0.07 (−2.01)* | −0.14 (−3.57)* |
| TAX | 0.000102 (3.28)*** | 0.0003 (1.29) | 6.3E-07 (5.51) | 0.09 (5.53)* | 0.008 (3.91) |
| TRS | −5.36E-06 (−2.38)*** | −0.0004 (−0.24) | −0.02 (−1.76)*** | | −0.0002 (−1.12) |
| FL | 0.0001 (−0.52) | −0.009 (−5.54)* | −0.01 (−0.33) | −0.4 (−4.96)* | −0.03 (−2.87)*** |
| CONST | 1.091 (1.51) | 0.879 (0.61) | 0.945 (14.25) | 1.3 (8.7) | 0.921 (15.34)* |
| $R^2$ | 0.27 | 0.34 | 0.5 | 0.4 | 0.39 |
| Observation | 72 | 72 | 72 | 72 | 72 |